# Configuration Interaction Study of the $^3P$ Ground State of the Carbon Atom

María Belén Ruiz* and Robert Tröger

Department of Theoretical Chemistry
Friedrich-Alexander-University Erlangen-Nürnberg
Egerlandstraße 3, 91054 Erlangen, Germany

## Abstract

Configuration Interaction (CI) calculations on the ground state of the C atom are carried out using a small basis set of Slater orbitals [7s6p5d4f3g]. The configurations are selected according to their contribution to the total energy. One set of exponents is optimized for the whole expansion. Using some computational techniques to increase efficiency, our computer program is able to perform partially-parallelized runs of 1000 configuration term functions within a few minutes. With the optimized computer programme we were able to test a large number of configuration types and chose the most important ones. The energy of the $^3P$ ground state of carbon atom with a wave function of angular momentum L=1 and $M_L$=0 and spin eigenfunction with S=1 and $M_S$=0 leads to -37.83526523 h, which is millihartree accurate. We discuss the state of the art in the determination of the ground state of the carbon atom and give an outlook about the complex spectra of this atom and its low-lying states.

**Keywords:** Carbon atom; Configuration Interaction; Slater orbitals; Ground state

*Corresponding author: e-mail address: maria.belen.ruiz@fau.de

# 1. Introduction

The spectrum of the isolated carbon atom is the most complex one among the light atoms. The ground state of carbon atom is a triplet $^3P$ state and its low-lying excited states are singlet $^1D$, $^1S$ and $^1P$ states, more stable than the corresponding triplet excited ones $^3D$ and $^3S$, against the Hund's rule of maximal multiplicity. While a quintet spin $^5S$ state is the third excited state, no other quintet is stable as a bound state of the carbon atom. In this work we study the $^3P$ ground state of carbon atom.

The first calculations on the $^3P$ ground state of carbon atom were carried out by Boys [1], Clementi [2] and Bagus [3] using small basis sets und by Schaefer et al. [4] with a larger basis. The short superposition of configurations used by Weiss [5] with only 40 terms and a basis set of n=5 or [5s4p3d2f] basis serves as a guide to start the calculations on the ground state. The first accurate Configuration Interaction (CI) calculations on the ground state of carbon atom where performed by A. Bunge [6] and posterior works of A. Bunge and C.F. Bunge in the early 70s [7,8]. A few years later Sasaki and Yoshimine [9] carried out very sophisticated CI calculations with the ALCHEMY computer program of the theoretical chemistry group at IBM in San Jose. They used natural orbitals of s, p, d, f, g, h and i symmetry formed with large numbers of atomic orbitals (in total 35) and large numbers of configurations. They obtained the best CI energy to date. This energy has only been improved recently by the use of the method of Exponential Correlated Gaussians (ECG) [10]. Other accurate approaches used in atoms are Quantum Monte Carlo [11] and Multi-configurational explicit correlated wave function [12]. A larger number of low-lying excited states of the carbon atom have been calculated using Variational and Diffusion Monte Carlo approaches [13].

In this work we show a simple methodology to perform CI calculations obtaining a good total energy using very little computer time. We show how to construct the configurations and select them and how to find the most important ones with respect to the energy. The Section 'Calculations' below contains the wave function expansion written in detail. Finally, we compare the obtained energy for the ground state with the results obtained by other authors and methods.

# 2. Configuration Interaction

The method of Configuration Interaction (CI) proposed by Löwdin [14] in 1955, the so-called 'Father of Quantum Chemistry', is a variational systematic method to calculate straightforwardly the electron correlation energy in atoms and molecules. In a recent volume celebrating the 100[th] anniversary of the birth of Per-Olov Löwdin, one of his students, Carlos F. Bunge, dedicated a chapter to him [15]. Another chapter about the state of the art in highly accurate CI calculations on atoms and molecules [16] is also recommended.

The CI method endeavours to recover rapidly a great part of the correlation energy but the method converges extremely slowly for obtaining highly accurate energy values. This is due to the shortcomings of the CI wave-function which form does not fulfil the electronic cusp-condition. The CI wave function contains excitations like p(1)p(2), d(1)d(2), which are proportional to $r_{12}^2$, $r_{12}^4$, … $r_{12}^{2n}$ terms, but not to a linear $r_{12}$ term, which is required by the cusp. Physically, the inter-electronic coordinate $r_{ij}$ represents the situation when the distance between two electrons gets close to zero. Therefore the CI wave function needs a huge number of Slater determinants to converge. Conversely, methods including explicitly $r_{ij}$ are the explicit correlated methods, introduced by the pioneering work of Hylleraas in 1929 [17], they converge faster to the exact solution of the Schrödinger equation. Nevertheless, in this work we employ the CI method as a first step for further Hylleraas-Configuration Interaction (Hy-CI) calculations on the ground state of carbon atom. For more information on the Hy-CI method see i.e. Refs. [18,19].

The CI wave function is constructed as a linear combination of N$_{conf}$ configurations, where the coefficients Cp are determined variationally by solving the eigenvalue problem which follows from the Schrödinger equation:

$$\Psi = \sum_{p=1}^{N_{conf}} C_p \Phi_p$$

(1)

The configurations $\Phi_p$ are symmetry adapted since they are eigenfunctions of the square of the angular momentum operator $\hat{L}^2$. Every configuration is then a linear combination of function terms and these ones are a linear combination of Slater determinants:

$$\Phi_p = \hat{O}(\hat{L}^2)\hat{\mathcal{A}}\psi_p \chi$$

(2)

In this work, the symmetry adapted configurations are constructed 'a priori', so that they are eigenfunctions of $\hat{L}^2$. Another possibility would be the posterior projection of the configurations over the proper space, as indicated in Eq. (2) by the projection operator $\hat{O}(\hat{L}^2)$. The Slater determinants are constructed with the help of the anti-symmetrisation operator $\hat{\mathcal{A}}$, which acts not only over the spatial orbitals but also over the spin function part.

The configurations are also eigenfunctions of the square of the spin operator $\hat{S}^2$. $\chi$ is a spin eigenfunction. In the case of the $^3$P ground state of carbon atom, a triplet state with S = 1, we have chosen for convenience a spin eigenfunction with M$_S$ = 0:

$$\chi = (\alpha\beta - \beta\alpha)(\alpha\beta - \beta\alpha)(\alpha\beta + \beta\alpha).$$

(3)

This spin function differs only by a sign from the singlet one S=1 for six electrons, and in this way the computer program can be used for singlet or triplet states only by changing one sign. There are more spin eigenfunctions for S=1, namely the ones with $M_S$=1 and $M_S$=-1, but they are all degenerate with respect to the energy. This means, it is indifferent which one we use and since they are orthogonal, it is sufficient to use only one of them.

As discussed in the case of the Li-atom in Ref. [20] and calculations of the Be-atom [24] it is also sufficient to consider only one spin-function. The spatial part of the basis-functions consists of Hartree products of Slater Type Orbitals (STOs):

$$\psi_p = \prod_{k=1}^{n_e} \phi(r_k, \theta_k, \varphi_k)$$

(4)

Every Slater orbital $\phi$ is represented by only one orbital s, p, d, f, g or i, therefore this is named the minimal basis set. The un-normalised STOs are defined:

$$\phi(r, \theta, \varphi) = r^{n-1} e^{-\alpha r} Y_l^m(\theta, \varphi)$$

(5)

where $Y_l^m$ are the spherical harmonics [20].

The Schrödinger equation to be solved is:

$$\widehat{H}\Psi = E\Psi$$

(6)

The atomic Hamiltonian for a fixed nucleus written in Hylleraas coordinates, see Ref. [23], for a CI wave function reduces effectively to:

$$\widehat{H} = -\frac{1}{2}\sum_{i=1}^{n_e}\frac{\partial^2}{\partial r_i^2} - \sum_{i=1}^{n_e}\frac{1}{r_i}\frac{\partial}{\partial r_i} - \sum_{i=1}^{n_e}\frac{Z}{r_i} + \sum_{i<j}^{n_e}\frac{1}{r_{ij}} + \sum_{i=1}^{n_e}\frac{1}{r_i^2}l_i(l_i + 1)$$

(7)

where $l_i$ are the angular quantum numbers of the orbitals or spherical harmonics. From the variational principle we have to solve the following matrix eigenvalue problem:

$$(\mathbf{H} - E\mathbf{S})\mathbf{C} = 0$$

(8)

with the matrix elements of the Hamiltonian **H** and overlap **S** matrices are:

$$\mathbf{H}_{kl} = \int \Phi_k \widehat{H} \Phi_l \, \partial\tau$$

$$\mathbf{S}_{kl} = \int \Phi_k \Phi_l \, \partial\tau$$

(9)

The matrix elements are sums of one- and two-electron integrals. Their expressions and calculation have been described in previous papers, see [24,25].

## 3. Selection of Configurations

In this work we use un-normalised non-orthogonal orbitals STOs. Our basis set can be called minimal basis, since for every electron a single STO is employed. This basis is [7s6p5d4f3g]. This is a difference with respect the natural orbitals employed in the works of Bunge [7] and Sasaki and Yoshimine [9], where an atomic orbital is a linear combination of basis orbitals and in addition they are optimized. In this work we have the restriction of same orbital exponent per pair of electrons. Later this issue will be further discussed.

The ground state leading configuration of the C-atom is $1s^2\,2s^2\,2p^2$, in our nomenclature ssspp (i.e. s(1)s(2)s(3)s(4)p(5)p(6)). The other configurations with large contributions to the energy for the P-symmetry (L=1) six-electron state are, ordered by decreasing energy contribution, ssppd, sspppp, sssdd, sppdd, pppppp, sssddd, ssssff, ssspdf, ssssgg, ssspfg, ssffpp, and ssggpp. The permutations of these configurations were also included. For instance, the configuration type ssspp, with the restriction of same exponent per electron pair, includes the configurations ssspsp, spspss, spssp, ppssss and sppss. Note that configurations like ppsspp or ddsspp do have a great energy contribution, because the inner excitations resemble implicitly the term $r_{12}$. The same permutations have been considered for the other configurations.

Other configurations than these have been tried and sorted out because of their very low energy contribution. As pointed out by Bunge [7,9], we encountered also two sets of configurations with the correct symmetry L=1, those that interact with the ground state main configuration or HF configuration and some others which do not interact at all with the first configuration. These non-interacting configurations are: sssssp, ssspd, ssssdf, sssppp, among other possible ones and their permutations. After trying these configurations for low n, they were sorted out of the expansion.

The quantum number $M_L = 0$ was chosen, because for this case the same spin primitives except for a sign than in the singlet state are required. We systematically selected the CI configurations according to their energy contribution. This was done by calculations on blocks constructed for all possible configurations. In these blocks all excitations are included from single, double, … up to the maximal number of excitations, in this case sextuple, whereas seldom sextuple excited configurations showed even a moderate contribution. The eigenvalue equation was diagonalized upon each addition of a configuration. In this manner, the contribution of every single configuration and of each block of a given type to the total energy was evaluated. If the energy difference was less than the threshold $|E_{i-1}-E_i| < 1.10^{-6}$ hartree (h), the new configuration was discarded. In this manner, all configurations were checked, leading to a relatively compact CI wave function. The procedure of selection of the configurations is similar to that described in our previous works, Refs. [23,24].

We have noticed that usually the larger the contribution of a configuration, the smaller the sum of the *l* quantum numbers of the employed orbitals $l_1 + l_2 + l_3 + l_4 + l_5 + l_6$ is; i.e. the contribution of the configuration sssspp > ssppppp for a P-state. Among the many possibilities to construct configurations of these symmetries, energetically important configurations were proven to be those with an inner S-shell and the outer pair of electrons with the P symmetry, for example ppsspp and ddsspp. Other important configurations for the ground state of carbon atom are of the form: ppd, pdf and pfg and all their permutations.

Note that there are more possible 'degenerate L-eigenfunction' solutions with a larger number of Slater determinants. Specifically, these are degenerate with respect to the quantum numbers L and M, but with possible different energy contribution, i.e. non-degenerate with respect to the energy [7]. Although the inclusion of various degenerate configurations has been shown to improve the energy of the state, such a contribution is very small. This becomes important for very accurate CI calculations, as reported e.g. by Bunge [16]. In our work, we have constructed the configurations as linear combinations of all different terms occurring in the different degenerate configuration, see Table 1.

In Table 1 a list of two- and three-electron configurations of $^1$S and $^3$P symmetry is shown. The total configuration is constructed as the product of these configurations. It is clear that the product of $S \times P = P$ leads to P symmetry. In this list sums, subtractions and coefficients are missing, because they all represent a configuration term and the corresponding coefficient is determined variationally during the calculation. Only the constituents of the configurations are given. For example, the configuration ppsspp is the product of ($^1$S pp) ($^1$S ss) ($^3$P pp). Other configurations not appearing in Table 1 have been already given in previous papers and are here omitted, see [20,23].

**Table 1: Configuration terms needed for the construction of two- and three-electron configurations with $M_L$=0.**

| Configuration | Configuration terms |
|---|---|
| ($^1$S pp) | $p_0 p_0$, $p_1 p_{-1}$, $p_{-1} p_1$ |
| ($^1$S dd) | $d_0 d_0$, $d_1 d_{-1}$, $d_{-1} d_1$, $d_2 d_{-2}$, $d_{-2} d_2$ |
| ($^1$S ff) | $f_0 f_0$, $f_1 f_{-1}$, $f_{-1} f_1$, $f_2 f_{-2}$, $f_{-2} f_2$, $f_3 f_{-3}$, $f_{-3} f_3$ |
| ($^1$S gg) | $g_0 g_0$, $g_1 g_{-1}$, $g_{-1} g_1$, $g_2 g_{-2}$, $g_{-2} g_2$, $g_3 g_{-3}$, $g_{-3} g_3$, $g_4 g_{-4}$, $g_{-4} g_4$ |
| ($^3$P pp) | $p_1 p_{-1}$, $p_{-1} p_1$ |
| ($^3$P dd) | $d_1 d_{-1}$, $d_{-1} d_1$, $d_2 d_{-2}$, $d_{-2} d_2$ |
| ($^3$P ff) | $f_1 f_{-1}$, $f_{-1} f_1$, $f_2 f_{-2}$, $f_{-2} f_2$, $f_3 f_{-3}$, $f_{-3} f_3$ |
| ($^3$P ff) | $g_1 g_{-1}$, $g_{-1} g_1$, $g_2 g_{-2}$, $g_{-2} g_2$, $g_3 g_{-3}$, $g_{-3} g_3$, $g_4 g_{-4}$, $g_{-4} g_4$ |
| ($^3$P ppd) | $p_1 p_1 d_{-2}$, $p_{-1} p_{-1} d_2$, $p_1 p_{-1} d_0$, $p_{-1} p_1 d_0$, $p_0 p_1 d_{-1}$, $p_0 p_{-1} d_1$, $p_1 p_0 d_{-1}$, $p_{-1} p_0 d_1$ |
| ($^3$P pdf) | $p_1 d_2 f_{-3}$, $p_{-1} d_{-2} f_3$, $p_0 d_2 f_{-2}$, $p_0 d_{-2} f_2$, $p_1 d_1 f_{-2}$, $p_{-1} d_{-1} f_2$, $p_1 d_{-2} f_1$, $p_{-1} d_2 f_{-1}$, $p_1 d_{-1} f_0$, $p_{-1} d_1 f_0$, $p_0 d_1 f_{-1}$, $p_0 d_{-1} f_1$, $p_1 d_0 f_{-1}$, $p_{-1} d_0 f_1$ |
| ($^3$P pfg) | $p_1 f_3 g_{-4}$, $p_{-1} f_{-3} g_4$, $p_1 f_{-3} g_2$, $p_{-1} f_3 g_{-2}$, $p_1 f_2 g_{-3}$, $p_{-1} f_{-2} g_3$, $p_0 f_3 g_{-3}$, $p_0 f_{-3} g_3$, $p_0 f_2 g_{-2}$, $p_0 f_{-2} g_2$, $p_1 f_1 g_{-2}$, $p_{-1} f_{-1} g_2$, $p_1 f_{-2} g_1$, $p_{-1} f_2 g_{-1}$, $p_1 f_{-1} g_0$, $p_{-1} f_1 g_0$, $p_0 f_1 g_{-1}$, $p_0 f_{-1} g_1$, $p_1 f_0 g_{-1}$, $p_{-1} f_0 g_1$ |
| ($^3$P ddd) | $d_2 d_{-2} d_0$, $d_{-2} d_2 d_0$, $d_0 d_2 d_{-2}$, $d_0 d_{-2} d_2$, $d_2 d_0 d_{-2}$, $d_{-2} d_0 d_2$, $d_2 d_{-1} d_{-1}$, $d_{-2} d_1 d_1$, $d_1 d_1 d_{-2}$, $d_{-1} d_{-1} d_2$, $d_1 d_{-2} d_1$, $d_{-1} d_2 d_{-1}$, $d_1 d_{-1} d_0$, $d_{-1} d_1 d_0$, $d_0 d_1 d_{-1}$, $d_0 d_{-1} d_1$, $d_1 d_0 d_{-1}$, $d_{-1} d_0 d_1$ |

## 4. Calculations

We have written a CI computer program for six-electron atomic systems in Fortran 95. Numerical calculations have been conducted with double precision arithmetic. The program is an extension of our four- and five-electron CI programs. The four-electron computer program has been thoroughly checked by comparing results of our numerical calculations with the results by Sims and Hagstrom for the Be atom [21]. In these calculations, we obtained complete agreement. And the five-electron CI program has been tested by comparing calculations with the ones of Bunge and Froese Fischer as reported in Ref. [25].

In this work we start with a full-CI wave function for a basis $n=3$ (or [3s2p1d]) testing all possible configurations and retaining all which contribute more than $1.10^{-6}$ h, which at the beginning of the calculation are almost all of them. First for a relatively short expansion the orbital exponents are optimized. We use the same technique than in previous calculations of the Be and Li atoms [20,23,24]. A set of three exponents is used (one for the K-shell, other for the electrons of the inner L-shell and other for the pair of outer electrons of the L-shell), and kept equal for all configurations. This technique accelerates computations, while still producing sufficiently accurate wave functions to determine the bound state properties. During the calculations and optimizations we use the virial theorem to control the quality of the wave function and guide the numerical optimization:

$$\sigma = -\frac{\langle V \rangle}{\langle T \rangle}$$

(10)

The following step is to add all the possible configurations with $n=4$. After this the orbital exponents are optimized once more and kept equal for the rest of the calculations. The wave function is very compact, it contains many different configurations and linear dependences produced by very similar configurations are avoided. This means that it is computationally favourable to vary the configurations within a quantum number than to make a calculation with one configuration for all quantum numbers. We have added in this work configurations up to $n=6$. At the end it is difficult to find configurations which achieve to add something to the energy. It is then clear that the contribution of a configuration depends on the order where it has been added. Nevertheless, this simple method is very stable and helps to retain the most important configurations.

In Table 2 the truncated CI wave function expansion is shown. The configurations with larger contribution are the ground state configuration sssspp, the open shell configuration ssspsp, the sssppd and sssdpp $^3$P configurations, followed by the configuration with inner $^1$S excitations ppsspp, afterwards we find the sspppp, which shows the quasi-degeneration of the *s* and *p* orbitals of the L-shell; the configurations ssssdd are contributed less as would be expected by the low $l_i$ quantum numbers, and higher excitations sspdpd, ddsspp and ssffpp follow among others.

**Table 2: Selected CI wave function expansion for the $^3$P ground state of carbon atom. N$_{conf}$ is the number of symmetry adapted configurations and *n* the basis set, i.e. n=3 means [3s2p1d].**

| Conf. | n | N$_{conf}$ | N$_{conf,tot}$ | Energy (h) | - Diff. ($\mu h$) | virial |
|---|---|---|---|---|---|---|
| sssspp | 6 | 181 | 181 | -37.72179917 | 0 | 2.00006 |
| ssspsp | 6 | 100 | 281 | -37.72756352 | 5764 | 2.00003 |
| ssppss | 3 | 1 | 282 | -37.72756615 | 2 | 2.00003 |
| spsssp | 5 | 44 | 326 | -37.72856844 | 1002 | 1.99999 |
| spspss | 4 | 6 | 332 | -37.72859600 | 27 | 1.99999 |
| sspppp | 6 | 50 | 382 | -37.75059877 | 22002 | 1.99999 |
| spsppp | 5 | 56 | 438 | -37.75377264 | 3173 | 1.99999 |
| spppsp | 3 | 1 | 439 | -37.75377312 | 0 | 1.99999 |
| ppsspp | 7 | 78 | 517 | -37.77733780 | 23564 | 2.00002 |
| ppspsp | 5 | 14 | 531 | -37.77749030 | 152 | 2.00001 |
| pppppp | 5 | 6 | 537 | -37.77797403 | 483 | 2.00002 |
| sssppd | 5 | 61 | 598 | -37.80802592 | 30051 | 2.00000 |
| sssdpp | 5 | 41 | 639 | -37.81502320 | 6997 | 2.00000 |
| sspspd | 5 | 20 | 659 | -37.81515432 | 131 | 2.00002 |
| sspdps | 5 | 14 | 673 | -37.81523446 | 80 | 2.00003 |
| ssdspp | 5 | 2 | 675 | -37.81523913 | 4 | 2.00003 |
| spsspd | 6 | 18 | 693 | -37.81759021 | 2351 | 2.00002 |
| spsdps | 7 | 36 | 729 | -37.81823607 | 645 | 2.00002 |
| sppdss | 3 | 4 | 733 | -37.81827887 | 42 | 2.00001 |
| sdsspp | 4 | 9 | 742 | -37.81848443 | 205 | 2.00001 |
| sdspsp | 5 | 36 | 778 | -37.81856553 | 81 | 2.00001 |
| sdppss | 3 | 2 | 780 | -37.81856602 | 0 | 2.00001 |
| pdsssp | 5 | 31 | 811 | -37.81881301 | 246 | 2.00002 |
| pdspss | 5 | 19 | 830 | -37.81885580 | 42 | 2.00002 |
| ssppdd | 3 | 2 | 832 | -37.81910329 | 247 | 2.00001 |
| sspdpd | 5 | 21 | 853 | -37.82326956 | 4166 | 2.00006 |
| ssddpp | 7 | 14 | 867 | -37.82360092 | 331 | 2.00003 |

| | | | | | |
|---|---|---|---|---|---|
| spsdpd | 3 | 2 | 869 | -37.82361411 | 13 | 2.00003 |
| sppdsd | 4 | 2 | 871 | -37.82363361 | 19 | 2.00003 |
| spdpsd | 3 | 1 | 872 | -37.82363620 | 2 | 2.00003 |
| sdsdpp | 6 | 6 | 878 | -37.82392325 | 287 | 2.00002 |
| ppssdd | 3 | 3 | 881 | -37.82392335 | 0 | 2.00002 |
| ddsspp | 5 | 26 | 907 | -37.82630085 | 2377 | 2.00009 |
| ssssdd | 4 | 13 | 920 | -37.82913117 | 2830 | 2.00011 |
| sssdsd | 5 | 21 | 941 | -37.82951963 | 388 | 2.00009 |
| ppddpp | 3 | 1 | 942 | -37.82953897 | 19 | 2.00009 |
| sssddd | 4 | 3 | 945 | -37.82966403 | 125 | 2.00010 |
| ssddsd | 4 | 6 | 951 | -37.82971989 | 55 | 2.00009 |
| ssspdf | 5 | 7 | 958 | -37.83022935 | 509 | 2.00010 |
| sssdpf | 7 | 16 | 974 | -37.83205398 | 1824 | 2.00012 |
| sssfpd | 6 | 11 | 985 | -37.83303456 | 980 | 2.00013 |
| sssfdp | 7 | 4 | 989 | -37.83307198 | 37 | 2.00013 |
| sspdsf | 4 | 8 | 997 | -37.83309783 | 25 | 2.00013 |
| sspfsd | 5 | 10 | 1007 | -37.83316362 | 65 | 2.00013 |
| ssdfsp | 6 | 20 | 1027 | -37.83350195 | 338 | 2.00010 |
| dfsssp | 4 | 4 | 1031 | -37.83365871 | 156 | 2.00010 |
| ssptpf | 5 | 4 | 1035 | -37.83385580 | 197 | 2.00009 |
| ssssff | 5 | 5 | 1040 | -37.83411703 | 261 | 2.00010 |
| sssfsf | 5 | 9 | 1049 | -37.83416895 | 51 | 2.00009 |
| ssffpp | 5 | 4 | 1053 | -37.83463042 | 461 | 2.00010 |
| ffsspp | 5 | 4 | 1057 | -37.83474788 | 117 | 2.00011 |
| ssspfg | 5 | 1 | 1058 | -37.83475897 | 11 | 2.00011 |
| sssfpg | 7 | 9 | 1067 | -37.83500422 | 245 | 2.00012 |
| sssgpf | 6 | 5 | 1072 | -37.83512518 | 120 | 2.00012 |
| ssfgsp | 5 | 1 | 1073 | -37.83513520 | 10 | 2.00012 |
| fgsssp | 5 | 1 | 1074 | -37.83516054 | 25 | 2.00013 |
| ssggpp | 6 | 3 | 1077 | -37.83523933 | 78 | 2.00013 |
| ssssgg | 5 | 1 | 1078 | -37.83524671 | 7 | 2.00013 |
| sssgsg | 6 | 2 | 1080 | -37.83526523 | 18 | 2.00013 |

The orbitals exponents are $\alpha = \alpha' = 6.39437109$, $\beta = \beta' = 1.78356693$, $\gamma = 1.68267232$ for all the expansion. The unit $1\mu h = 1 \times 10^{-6} a.u$

If we order the energy for the basis set *n*, we encounter the angular energy limits, assuming the wave function is saturated for these limits. The contribution to the total energy decreases with increasing *n*, see Table 3. That the contribution of *n=7* is still not close to zero indicates that the wave function has not fully converged. We would need to add more terms with *n=8* and higher. We estimate that the contribution to the energy of the missing *g,* as well *h, i* and higher orbital configurations is about 4200 $\mu h$ (microhartree).

Table 3: Angular energy limits for the ³P ground state of carbon atom.

| n | Basis set | N$_{conf}$ | N$_{conf,tot}$ | Energy (a.u.) | - Diff. ($\mu h$) | virial |
|---|---|---|---|---|---|---|
| n = 2 | [2s1p] | 66 | 66 | -37.67686631 | 7981595 | 1.98191 |
| n = 3 | [3s2p1d] | 332 | 398 | -37.81179154 | 134925 | 1.99911 |
| n = 4 | [4s3p2d1f] | 358 | 756 | -37.83030344 | 18511 | 2.00021 |
| n = 5 | [5s4p3d2f1g] | 288 | 1044 | -37.83458668 | 4283 | 2.00019 |
| n = 6 | [6s5p4d3f2g] | 27 | 1071 | -37.83510448 | 517 | 2.00015 |
| n = 7 | [7s6p5d4f3g] | 9 | 1080 | -37.83526520 | 160 | 2.00013 |

*n* is the basis set indicated by the highest principal quantum number.

Our energy result is very close to that of Bunge using similar basis set. In that work Bunge used about 5000 Slater determinants. In our work the number of configurations is larger than in other works, because we use a basis of one orbital for one electron. The calculation of Sasaki and Yoshimine is highly elaborated using up to i-orbitals and 900 configurations. We could estimate that they used more than 5000 of our type configurations. They have managed to recover a great amount of correlation energy by using the CI method. Their calculation can be considered as the CI benchmark for the carbon atom. We can observe than the calculation using the ECG method, which includes all inter-electronic coordinates, is close in value to Sasaki and Yoshimine calculation.

Table 4: Comparison of variational upper bonds to the energy of the 3P ground state of carbon atom calculated by different methods.

| Method | Orb. | Authors | Year | Ref. | Basis Set | N$_{confs}$ | Energy (h) |
|---|---|---|---|---|---|---|---|
| CI | STO | Weiss | 1967 | [5] | [4s3p2d2f] | 40 | -37.77888 |
| VMC | | Sarsa et al. | 2016 | [12] | | | -37.81537 |
| FCI | STO | Bunge et al. | 1970 | [8] | [7s6p4d3f] | 234 | -37.83378 |
| CI | STO | This work | 2017 | | [7s6p5d4f3g] | 1080 | -37.835265 |
| DMC | | Maldonado et al. | 2011 | [13] | | | -37.83544(9) |
| CI(SDTQ) | STO | Sasaki et al. | 1974 | [9] | [10s9p8d8f6g4h2i] | 993 | -37.8393 |
| ECG | GTO | Sharkey et al. | 2010 | [10] | | 500 | -37.84012879 |
| MC | | Seth et al. | 2011 | [11] | | | -37.84446 |
| Estimated exact | | Chakravorty et al. | 1993 | [27] | | | -37.8450 |

# 5. Conclusions and Perspectives

We have calculated the ground state of carbon atom with millihartree accuracy and modest computational effort. We have determined the important configurations of the ground states. The relevant configurations together with the leading one sssspp are sssppd and followed by sspppp and ssssdd, showing that the excitation pdd is more important than the quasi-degeneration of the s and p levels as it was the case in the Be atom. The weight of the configurations ppsspp, ddsspp, ssffpp can be interpreted as description of the cusp $r_{12}$ with inner singlet $^1$S excitations pp and dd. Finally, the configuration ssspsp adds the orbital splitting of the orbitals, which was missing in sssspp due to the imposed double occupancy.

In the same way, the low-lying states of carbon atom can be determined. The spectrum of the low-lying excited states of carbon atom looks very complex. Maldonado et al. calculated a large number of low-lying excited states employing excitation energies obtained with Variational and Diffusion Monte Carlo approaches [13]. Their spectrum serves as a guide for future CI calculations. It agrees very well with our estimations calculated adding to the estimated exact energy of the ground state -37.8450 h the excitation energies (in hartree) of the levels, which are collected in the Atomic Data Base of NIST [26]. These estimated energies of bound states of the carbon atom are given in Table 5 and compared with the energy values obtained from Monte Carlo excitation energies. The results show that carbon atom represents a vast area of research.

**Table 5: Estimated total energies of the low-lying states of C atom.**

| Configuration | State | Energy (cm$^{-1}$) [26] | Estimated exp. (h) | DMC Ref. [13] |
|---|---|---|---|---|
| $2s^2\,2p^2$ | $^3$P | 0.0 | -37.8450 | -37.83544 |
| $2s^2\,2p^2$ | $^1$D | 10193 | -37.7985 573 | -37.78966(6) |
| $2s^2\,2p^2$ | $^1$S | 21648 | -37.7463 645 | -37.73830(5) |
| $2s\,2p^3$ | $^5$S | 33735 | -37.6912 92 | -37.69026(3) |
| $2s^2\,2p3s$ | $2\,^3$P | 60333 | -37.5701 026 | -37.56103(10) |
| $2s^2\,2p3s$ | $^1$P | 61981 | -37.5625 938 | -37. 55352(6) |
| $2s\,2p^3$ | $^3$D | 64091 | -37.5529 799 | -37. 54374(8) |
| $2s^2\,2p3p$ | $2\,^1$P | 68856 | -37.5312 690 | -37. 52335(7) |
| $2s^2\,2p3p$ | $2\,^3$D | 69689 | -37.5274735 | -37. 51974(13) |
| $2s^2\,2p3p$ | $^3$S | 70743 | -37.5226712 | -37.51523(5) |
| $2s^2\,2p3p$ | $3\,^3$P | 71352 | -37.5198964 | -37.49697(5)* |
| $2s^2\,2p3p$ | $2\,^1$D | 72611 | -37.5141599 | -37.49096(3)* |
| $2s^2\,2p3p$ | $2\,^1$S | 73976 | -37.5079405 | -37.47429(3)* |
| $2s\,2p^3$ | $4\,^3$P | 75254 | -37.5021175 | -37.46760(10)* |
| $2s^2\,2p3d$ | $3\,^1$D | 77680 | -37.4910639 | -37.48274(8) |
| $2s^2\,2p4s$ | $5\,^3$P | 78148 | -37.4889315 | -37.46084(3)* |
| $2s^2\,2p3d$ | $^3$F | 78199 | -37.4886991 | -37.48097(13) |
| Ioniz. | | 90820 | -37.4311936 | |

For the energy conversion we have used 1h = 219474.6305 cm$^{-1}$. The ground state energy is taken as -37.8450 h. With $^*$ are marked energies calculated with the Variational Monte Carlo approach [13].

## Acknowledgments

One of us (M.B. Ruiz) would like to thank Carlos Bunge for very interesting discussions on the construction of configurations in the CI method and to James S. Sims for providing integral values of two-electron integrals to test our codes and fruitful discussions about the general treatment of angular momentum in the wave functions. Both of us are very thankful to the Editor of this volume Philip E. Hoggan for the kind invitation to participate in it and for reading the manuscript.